%% file: ms.tex
\documentclass[preprintnumbers,amsmath,amssymb,prd,superscriptaddress,notitlepage,longbibliography]{revtex4-1}
\pdfoutput=1

\usepackage{graphicx}% Include figure files
\usepackage{dcolumn}% Align table columns on decimal point
\usepackage{bm}% bold math
\usepackage{wasysym}
\usepackage{threeparttable}

\input{mydefinitions}

\newcommand{\vQp}{\vQ^\prime}

\begin{document}

\title{Improved renormalization group computation of likelihood functions for 
cosmological data sets}

\author{Patrick McDonald}
\email{PVMcDonald@lbl.gov}
\affiliation{Lawrence Berkeley National Laboratory, One Cyclotron Road,
Berkeley, CA 94720, USA}

\date{\today}

\begin{abstract}

Evaluation of likelihood functions for cosmological large scale structure data 
sets (including CMB, galaxy redshift surveys, etc.) naturally involves 
marginalization, i.e., integration, over an unknown underlying random signal 
field.
Recently, I showed how a renormalization group method can be used 
to carry out this integration efficiently by first integrating out the 
smallest scale structure, i.e., localized structure on the scale of 
differences between nearby data cells, then combining adjacent cells in a
coarse graining step, then repeating this process over and over until all 
scales have been integrated. Here I extend the formulation in several ways in 
order to reduce the prefactor on the method's linear scaling with data set 
size. The key improvement is showing how to integrate out
the difference between specific adjacent cells before summing them in the 
coarse graining 
step, compared to the original formulation in which small-scale fluctuations
were
integrated more generally. I suggest some other improvements in details of 
the scheme, including showing how to perform the integration 
around a maximum likelihood estimate for the underlying random field. In the 
end, an accurate likelihood computation for a million-cell Gaussian test data 
set runs in two minutes on my laptop, with room for further optimization and 
straightforward parallelization. 

\end{abstract}

\maketitle

\section{Introduction}

\cite{2019PhRvD..99d3538M} presented a new method to evaluate large scale
structure
likelihood functions, inspired by renormalization group (RG) ideas from 
quantum field theory \cite[e.g.,][]{1974PhR....12...75W,2008mqft.book.....B}.
This paper is a followup to that one, so some of the pedagogical discussion 
and derivations there will not be repeated here.
To recap the basics: the fact that structure in the 
Universe starts as an almost perfectly 
Gaussian random field and evolves in a computable way on the largest
scales
\cite[e.g.,][]{1993ppc..book.....P,2009JCAP...08..020M,2018PhRvD..97b3508M} 
suggests a statistically rigorous
first-principles likelihood analysis can be used to extract information on
cosmological models from observational data sets
\cite[e.g.,][]{1998PhRvD..57.2117B,2004PhRvD..70h3511W,
2008MNRAS.389..497K,2009PhRvD..80j5005E,2019AnP...53100127E}.
Generally, we have a data vector $\vo$, some relatively small number of 
global cosmological parameters we want to measure, $\vtheta$, and a
random field we'd like to marginalize over, $\vphi$. ($\vphi$ could be a
variety of different things, depending on the data set and theoretical 
setup, e.g., the underlying true temperature field for CMB,
the linear
regime density and/or potential fields for a galaxy redshift survey 
modeled by traditional perturbation theory, the evolving displacement
field in the functional integral formulation of \cite{2018PhRvD..97b3508M},
etc.) 
Starting with Bayes' rule $L(\vtheta,\vphi|\vo)L(\vo)=
L(\vo|\vtheta,\vphi)L(\vphi,\vtheta)=
L(\vo|\vtheta,\vphi)L(\vphi|\vtheta)L(\vtheta)$
we obtain
\be
L(\vtheta|\vo)=
\int d\vphi~ L(\vtheta,\vphi|\vo)=
\int d\vphi~ L(\vo|\vtheta,\vphi)L(\vphi|\vtheta)~,
\ee
where I have dropped $L(\vo)$ which has no parameter dependence and 
the prior $L(\vtheta)$ which plays no role in this discussion because it can
be pulled out of the integral. 
I have highlighted the usual cosmological form where some of the cosmological
parameters determine a prior on the signal field, $L(\vphi|\vtheta)$, and
then there is some likelihood for the observable given $\vtheta$ and $\vphi$, 
$L(\vo|\vtheta,\vphi)$. It is this $\vphi$ integral that we need to 
carry out. 
Generally, we can take at least part of
$L(\vphi|\vtheta)$, $L_G(\vphi|\vtheta)$, to be Gaussian,
defined by its covariance,
$\vP(\vtheta)$.
In this case we have 
\be
L(\vtheta|\vo)=
\int d\vphi~ e^{-\frachalf \vphi^t \vP^{-1} \vphi - \frachalf
{\rm Tr}\ln(2 \pi \vP)+\ln L_{\rm NG}(\vphi|\vtheta)
+\ln L(\vo|\vtheta,\vphi)}~,
\label{eq:basicL}
\ee
where I have used $\ln\det(\vP)={\Tr}\ln(\vP)$ and defined
$\ln L_{\rm NG}(\vphi|\vtheta)\equiv 
\ln L(\vphi|\vtheta)-\ln L_G(\vphi|\vtheta)$.
(Even for what we call non-Gaussian initial conditions
\cite[e.g.,][]{2008PhRvD..78l3519M,
2010JCAP...03..011B,
2010PhRvD..81f3530G,2011MNRAS.417L..79G,2014arXiv1412.4671A,
2018JCAP...01..010M},
the observable can often if not always be written as
a function of an underlying Gaussian random field, i.e., no $L_{\rm NG}$
needed, and in other scenarios
like \cite{2018PhRvD..97b3508M} where the natural $\vphi$ is not Gaussian,
there is still a natural Gaussian piece.)
Less generally but still often usefully (e.g., for primary CMB and 
large scale galaxy clustering ignoring primordial non-Gaussianity) we can take 
$\ln L_{\rm NG}=0$ and
$L(\vo|\vtheta,\vphi)$ to be Gaussian by assuming $\vo$ is linearly 
related to $\vphi$, i.e.,
$\vo = \vmu +\vR \vphi +\vep$ where $\vmu$ is the mean vector,
$\vR$ is a linear 
response matrix,
and $\vep$ is Gaussian observational noise with covariance
matrix $\vN$. Then we
have
\begin{eqnarray}
\label{eq:LGausswithmu}
L_{\rm Gaussian}(\vtheta| \vo)
&=&\int d\vphi ~e^{-\frac{1}{2}\vphi^t \vP^{-1}\vphi 
- \frachalf{\rm Tr}\ln(2 \pi \vP)-
\frac{1}{2}(\vo-\vmu-\vR\vphi)^t \vN^{-1}(\vo-\vmu-\vR\vphi)
- \frachalf {\rm Tr}\ln(2 \pi \vN)} 
 \\ \nonumber
&=& e^{-\frac{1}{2}(\vo-\vmu)^t \vC^{-1}(\vo-\vmu)
- \frachalf{\rm Tr}\ln(2 \pi \vC)}
\end{eqnarray}
where in the last line the integration has been carried out analytically, 
with $\vC\equiv \vN+\vR\vP\vR^t$.
Even this analytic integration does not really solve the Gaussian problem, 
however, as the time to
calculate
$\vC^{-1}$ and $\det(\vC)$ (or its derivatives) 
by brute force numerical linear algebra routines 
scales like $N^3$, where $N$ is the
size of the data set, which becomes prohibitively slow for large data sets.
The RG approach of \cite{2019PhRvD..99d3538M} addresses the Gaussian scenario
by doing the $\vphi$ integral in a different way that produces the result
directly as a number instead of these matrix expressions, 
and can also be applied to non-Gaussian scenarios.
Note that, as discussed in 
\cite{2019PhRvD..99d3538M}, the approach can also be used to directly compute 
derivatives
of $\ln L(\vtheta|\vo)$ with respect to $\vtheta$, not just the value at one 
choice of $\vtheta$, by passing the derivative inside the $\vphi$ integral
to produce a new integral. Traditional power spectrum estimation can be done
by taking $\vtheta$ to parameterize $\vP(\vtheta)$ by amplitudes in 
$k$ bands. 

In spite of the fact that fairly fast methods to evaluate 
at least the Gaussian likelihood [Eq. (\ref{eq:LGausswithmu})]
have existed for a long time
\cite[e.g.,][]{2003MNRAS.346..619P,2003NewA....8..581P,
2007PhRvD..76d3510S,2017JCAP...12..009S,2018JCAP...01..003F},
more often in practice data analysts compute summary statistics
not explicitly based on likelihood functions
\citep[e.g.,][]{2016A&A...594A..11P,2017MNRAS.464.3409B}, calibrating their
parameter dependence and covariance by computing the same statistics 
on mock data sets.
It is not
entirely clear why existing likelihood-based methods are not used more often, 
and in 
\cite{2019PhRvD..99d3538M} I was cautious about advocating immediate 
implementation of the RG approach.
One question was if the prefactor on the 
linear scaling of computation time with data set size for this method might
be so large as to make it significantly slower than others.
This paper demonstrates that this is not a significant obstacle. 
At two minutes to accurately compute the likelihood function for a million-cell 
Gaussian test data set, 
the method is as fast as any that takes more than a few 
well-preconditioned conjugate gradient maximum likelihood solutions for the 
same data set (i.e., as fast as any method I know of, barring the possibility
that my Julia implementation of conjugate gradient maximum likelihood is 
unfairly slow). 
The only reason not to implement this is if you believe the whole idea
of likelihood-based analysis is a distraction. 
That would not necessarily 
be an entirely unreasonable
position. E.g., if you believe that there is a lot of reliable cosmological
constraining power to be gained from the deeply non-linear regime, 
heuristic summary statistics/``machine learning," combined with 
exhaustive 
mocks/simulations is probably the only way to extract it. To me, however,
the likelihood+RG approach proposed here seems like an appealing path to 
large scale analysis, especially for incorporating weakly nonlinear 
information (e.g., without the need to explicitly estimate a bispectrum
and its covariance). 

This paper lays out a series of essentially technical improvements to the 
basic approach presented in \cite{2019PhRvD..99d3538M}. See that paper for
a derivation of the general RG equation and some more pedagogical discussion.
Some of that basics are explained in less detail here when they can be read
there.

\section{Revised formulation}

\subsection{Master RG equation}

Consider the general functional integral over some field $\vphi$,
\be
I\equiv \int d\vphi~ e^{-S(\vphi)}\equiv 
\int d\vphi~ e^{-\frachalf \vphi^t \vQ^{-1} \vphi- 
\frachalf {\rm Tr} \ln \left(2 \pi \vQ\right)
- S_I(\vphi)} ~.
\label{eq:IderiveRG}
\ee
The connection to our cosmological likelihood functions, 
Eq. (\ref{eq:basicL}), is obvious, but not necessary for this subsection.
Suppose that $\vQ \rightarrow 0$, i.e., $\vQ^{-1}$ goes to infinity 
(all its eigenvalues). In that limit the $\vQ$ part of $I$ becomes 
a representation of the delta function and it is clear that 
$I(\vQ\rightarrow 0) \rightarrow \exp[-S_I(0)]$, i.e., the integral can
be done trivially. 
Generally, however, $\vQ$ is not sufficiently small 
so if we want to do the integral this way we need to change
$\vQ$ to take it to zero. But we can't simply change $\vQ$ because that will
change the value of $I$, the integral we are trying to 
perform. If we want to 
change $\vQ$ while preserving $I$ we need to simultaneously change $S_I$. 
The renormalization group equation tells us how to do this.
Guided by, e.g., \cite{2008mqft.book.....B}, 
\cite{2019PhRvD..99d3538M} showed that
we can preserve the value of $I$ if the following differential equation
is satisfied:
\be
S_I^\prime =
\frachalf
\frac{\partial S_I}
{\partial \vphi^t}
\vQp
\frac{\partial S_I}
{\partial \vphi}
-\frachalf {\rm Tr}\left[
\vQp
\frac{\partial^2 S_I}
{\partial \vphi \partial \vphi^t}
\right] ~,
\label{eq:masterRG}
\ee
where we parameterize the
evolution by $\lambda$, i.e., $\vQ=\vQ(\lambda)$, $S_I=S_I(\lambda)$, and
the prime means derivative with respect to $\lambda$, where 
$\vQ(\lambda=0)$ and $S_I(\lambda=0)$ represent the original elements of
the integral.
(Note that, relative to Eq. (7) of \cite{2019PhRvD..99d3538M}, 
I have moved the 
normalization constant $\sN$ into $S_I$, after extracting ${\rm Tr} \ln \vQ$ 
from it to keep the integral unit normalized when $S_I=0$.) 
This formula is pure math, i.e., it assumes essentially nothing about 
$\vQ$, $\vQ^\prime$, and $S_I(\vphi)$.
Typically $\lambda$ will represent a length scale, where structure in 
$\vQ$ has already been erased on smaller scales, and $\vQ^\prime$ is doing
the job of erasing it on scale $\lambda$, but Eq. (\ref{eq:masterRG}) 
applies to any infinitesimal change in $\vQ$.

\subsection{Application to Gaussian cosmological data}

As in \cite{2019PhRvD..99d3538M}, I will demonstrate the calculation 
for a purely Gaussian example, i.e., $S_I(\vphi)$ at most quadratic in $\vphi$.
This is a special case only---Eq. (\ref{eq:masterRG}) applies for any 
$S_I(\vphi)$.
The likelihood function will be Eq. (\ref{eq:LGausswithmu}), except for 
simplicity I will set $\vmu=0$, i.e., I take
\be
L(\vtheta| \vo)=\int d\vphi ~\frac{
e^{-\frac{1}{2}\vphi^t \vP^{-1}\vphi -
\frac{1}{2}(\vo-\vR\vphi)^t \vN^{-1}(\vo-\vR\vphi)} }{
\sqrt{\det(2\pi \vP)
\det(2\pi \vN)}}~.
\label{eq:GaussL}
\ee
For the RG method to be efficient, the linear response matrix $\vR$ and
the observational noise $\vN$ cannot be completely arbitrary. 
Ideally $\vR$ should be fairly short range, e.g., a 
CMB telescope beam convolution
or redshift space cells in which we have counted galaxies. 
Similarly, $\vN$ should be short-range, e.g., diagonal for 
uncorrelated 
noise. The general approach can be adapted for special kinds of deviations
from short range $\vR$ or $\vN$, but I will assume they are 
short range here.
I generally assume the problem can be formulated to make 
$\vP$ translation invariant (i.e., diagonal in Fourier space), 
although slow evolution
in statistics can easily be accommodated.  
It is potentially useful to change integration variables to 
$\vdelta\equiv \vphi- \vphi_0$, where $\vphi_0$ is some constant field 
specified by hand. 
We plan to make
$\vphi_0$ the maximum likelihood field, but do not need to assume that.
Substituting this into Eq. (\ref{eq:GaussL}) and comparing to 
Eq. (\ref{eq:IderiveRG}), understanding that $\vphi$ 
in Eq. (\ref{eq:IderiveRG})
is a dummy variable so we can just as well replace it with $\vdelta$, 
we see that
the general integral $I$ in Eq. (\ref{eq:IderiveRG})
is equivalent to the
the Gaussian cosmological $L(\vtheta|\vo)$ if we define
\be
\vQ^{-1}(0) \equiv \vP^{-1}+\vA_\star ~.
\ee
and
\begin{eqnarray}
\label{eq:SI0}
S_I(0) &\equiv& 
\vphi_0^t \vP^{-1} \vdelta+
\frachalf \vphi_0^t \vP^{-1} \vphi_0+
\frachalf(\vo-\vR\vphi_0-\vR\vdelta)^t 
\vN^{-1}(\vo-\vR\vphi_0-\vR\vdelta) 
+\frachalf {\rm Tr} \ln \left(2 \pi \vN\right) 
\\ \nonumber
& & -\frachalf \vdelta^t \vA_\star \vdelta
+ 
\frachalf {\rm Tr} \ln \left(\vI+\vA_\star\vP \right)
\\ \nonumber
&=&
\frachalf\vdelta^t \left(\vR^t \vN^{-1}\vR-\vA_\star\right)\vdelta
-\left[\left(\vo-\vR\vphi_0\right)^t \vN^{-1}\vR -\vphi_0^t \vP^{-1}\right]
\vdelta \\ \nonumber
& & + \frachalf \vphi_0^t \vP^{-1} \vphi_0
+\frac{1}{2}\left(\vo-\vR\vphi_0\right)^t\vN^{-1}\left(\vo-\vR\vphi_0\right)
+ \frachalf {\rm Tr} \ln \left(2 \pi \vN\right) 
+\frachalf {\rm Tr} \ln \left(\vI+\vA_\star\vP \right)
~.
\end{eqnarray}
The reason for subtracting $\frachalf \vdelta^t \vA_\star \vdelta$
from $S_I(0)$ and adding it to the
$\vQ^{-1}$ term (adding zero overall, with $\vA_\star$ an as yet unspecified
matrix) will become clear below.

As in \cite{2019PhRvD..99d3538M}, the evolving Gaussian 
$S_I(\lambda)$ is represented numerically by the evolving coefficients
$\vA(\lambda)$, $\vb(\lambda)$, and $\sN(\lambda)$ of the general form
\be
S_I(\lambda) \equiv 
\frac{1}{2}\vdelta^t ~ \vA(\lambda)~ \vdelta 
-\vb^t(\lambda) ~\vdelta
+\sN(\lambda) ~.
\label{eq:generalS}
\ee
Comparison to Eq. (\ref{eq:SI0}) for $S_I(0)$ 
sets the initial conditions for $\vA$, $\vb$, and $\sN$:
\be
\vA(0)\equiv \vR^t \vN^{-1}\vR-\vA_\star
\ee
\be
\vb(0)\equiv \vR^t \vN^{-1} \left(\vo-\vR\vphi_0\right) -\vP^{-1}\vphi_0
\ee
and 
\be
\sN(0) \equiv 
 \frachalf \vphi_0^t \vP^{-1} \vphi_0+
\frac{1}{2}\left(\vo-\vR\vphi_0\right)^t\vN^{-1}
\left(\vo-\vR\phi_0\right)
+\frachalf {\rm Tr} \ln \left(2 \pi \vN\right)
+ \frachalf {\rm Tr} \ln \left(\vI+\vA_\star\vP \right)~.
\label{eq:sN0}
\ee
Plugging Eq. (\ref{eq:generalS}) into 
Eq. (\ref{eq:masterRG}) we find the flow equations for
$\vA$, $\vb$, and $\sN$:
\be
\vA^\prime= \vA \vQ^{\prime} \vA
\label{eq:Aprime}
\ee
\be
\vb^\prime=\vA \vQ^{\prime} \vb 
\label{eq:bprime}
\ee
\be
\sN^\prime= \frachalf \vb^t \vQ^{\prime}  \vb
 -\frachalf {\rm Tr}\left[\vA \vQ^{\prime}\right]~.
\label{eq:Nprime}
\ee

Note that if $\phi_0$ is the maximum likelihood field (for given values of
$\vP$, $\vR$, etc.), $\vb=\vb(0)= 0$. 
If the problem happened to be 
statistically homogeneous (translation invariant), we could set 
$\vA_\star=\vR^t \vN^{-1}\vR$ to make $\vA=\vA(0)=0$. 
In that case there would
be no evolution---$\sN(0)$ would simply be the answer. This is the point of
$\vA_\star$, i.e., if we choose it to be as close as possible to 
$\vR^t \vN^{-1}\vR$, 
we can reduce the RG evolution to be a minimal
correction due to statistical inhomogeneities. 
The limitation, i.e., why $\vA_\star$ generally can only approximate
$\vR^t \vN^{-1}\vR$, is that $\vA_\star$ must maintain the symmetries 
necessary to 
allow us to efficiently evaluate ${\rm Tr} \ln \left(\vI+\vA_\star\vP \right)$ 
in Eq. (\ref{eq:sN0}),
e.g., in Fourier space, to set the initial value of $\sN$.

In terms of these definitions, the result of formal analytic integration is
\be
L(\vtheta|\vo)=
e^{\frachalf\vb^t \left(\vQ^{-1}+\vA\right)^{-1}\vb -\sN-
\frachalf{\rm Tr}\ln\left(\vI + \vA\vQ\right)}~.
\label{eq:finalintegral}
\ee
We can use this formula once the components have been coarse-grained 
sufficiently to allow brute force linear algebra.
To be clear: if we plug $\vA(0)$, $\vb(0)$, $\sN(0)$, and $\vQ(0)$ into this
equation, it becomes precisely the analytic integration result in 
Eq. (\ref{eq:LGausswithmu}) (with $\vmu=0$). The difference is that as these 
quantities
evolve and are coarse grained their dimensions become smaller, 
with the result of the 
small-scale integration that has been performed stored in the simple
number $\sN$. See \cite{2019PhRvD..99d3538M} for more discussion.

\subsection{Integrating out the difference between adjacent cells 
\label{sec:pairdQ}}

In \cite{2019PhRvD..99d3538M} I used
\be
\vQ^{-1}(\lambda)=\vQ^{-1}(0)+\vK(\lambda)
\ee
where $\vK(\lambda\rightarrow \infty)\rightarrow \infty$ to suppress 
fluctuations.
I mentioned the potentially cleaner possibility 
\be
\vQ(\lambda)=\vQ(0)\vW(\lambda)
\ee
where $\vW(\lambda\rightarrow \infty)\rightarrow 0$, e.g., 
$W(k,\lambda) \equiv e^{-k^2\lambda^2}$. Either of these was envisioned to 
suppress fluctuations in a smooth, homogeneous way (i.e., with no explicit
connection to the data cell structure), starting from small scales 
to large. Once fluctuations were sufficiently suppressed on the scale of 
data cells, adjacent cells were combined, i.e., adjacent elements in $\vb$ 
and the corresponding $2\times 2$ block in $\vA$ were summed. 
This worked well enough, but the number of elements that I
needed to store in $\vA$, which determines the speed of computation,
seemed surprisingly large. 

Here I introduce a new possibility, to more explicitly integrate out the
fluctuations between pairs of cells that we are going to combine 
(see Appendix \ref{ap:altpairint} for an alternative version of this idea). 
Given covariance matrix $\vQ^1$ for some vector, we know that the covariance
for a new vector where each adjacent pair of elements is replaced by one 
element with its average, $\vQ^{2c}$, is simply given by the average of the 
appropriate $2\times 2$ blocks of $\vQ^1$, e.g., 
$Q^{2c}_{11} =\frac{1}{4}(Q^1_{11}+Q^1_{12}+Q^1_{21}+Q^1_{22})$, 
$Q^{2c}_{12} =\frac{1}{4}(Q^1_{13}+Q^1_{14}+Q^1_{23}+Q^1_{24})$, etc. 
This makes clear
that if we define $\vQ^\prime \propto \vQ^{2} -\vQ^1$, where
$\vQ^{2}$ is the matrix of equivalent dimension to $\vQ^1$ but with
the $2\times 2$ blocks that will be compressed to $\vQ^{2c}$ replaced by their
average (e.g., $Q^{2}_{11}=Q^{2}_{12}=Q^{2}_{21}=Q^{2}_{22}=Q^{2c}_{11}$), 
we can straightforwardly evolve Eq. (\ref{eq:masterRG}) from a starting 
$\vQ^1$ to ending $\vQ^{2}$, followed by a coarse graining combination of 
cells, and repeat.
Formally, for each iteration what we are doing is defining 
$\vQ(\lambda) = \vQ^1+\lambda(\vQ^2-\vQ^1)$ so that 
$\vQ^\prime \equiv d\vQ/d\lambda=\vQ^2-\vQ^1$, and solving the 
differential equation (\ref{eq:masterRG}) for $\lambda$ running from 0
[where $\vQ(\lambda=0)=\vQ^1$] to 1 [where $\vQ(\lambda=1)=\vQ^2$]. 

The obvious problem here is that generally $\vQ^2-\vQ^1$ is a dense matrix, 
which we can't have if the method is to be fast. The key to the RG approach
working is that elements of $\vQ^2-\vQ^1$ will generally be small very 
far off-diagonal, i.e., physically we do not expect the correlation at wide
separations to change much when the separation is changed by a small fractional
amount. To put it another way, we do not expect to need to use small cells
when measuring correlations at wide separations. 
This allows us to drop most elements of $\vQ^2-\vQ^1$, keeping it, and
$\vA$ as influenced by it, sparse. 
The closest thing to an exception to this ``no fine structure at
large separations'' rule that comes to mind is the
BAO feature---a relatively narrow bump at wide separation. Considering such
a thing, we observe that it is only necessary for $\vQ^\prime$ to remain 
sparse, not strictly near-diagonal, i.e., we can if necessary include a strip
of elements somewhere off-diagonal in $\vQ^\prime$, propagate this into 
$\vA$, etc., as long as there are not too many of these elements. 

Operationally, this program is surprisingly straightforward. 
I start by computing 
one full row of $\vQ(0)=\left(\vP^{-1}+\vA_\star\right)^{-1}$. 
This is basically just a standard 
computation of a correlation function given a power spectrum,
i.e., this matrix obeys
translation invariance by construction, so its elements are a function only 
of separation, inverses can be done in Fourier space, and one row is all that 
is necessary to capture the full
matrix. 
This $\vQ(0)$ becomes $\vQ^1$ described
above and I compute the first two rows of $\vQ^2$ (the $2\times 2$ 
block-averaged
matrix) directly from it. 
From this I compute the full sparse $\vQ^\prime$ including only elements
above some threshold. I define the threshold to be some fraction of the
maximum absolute value of $\vQ^\prime$, called $\epsilon_{\vQ^\prime}$,
i.e., I keep elements with 
$|Q_{ij}^\prime|>\epsilon_{\vQ^\prime} {\rm max}|\vQ^\prime|$. Note that this 
makes no assumption about the structure of $\vQ^\prime$, e.g., an off-diagonal
stripe due to something like BAO will be propagated if it passes the threshold.

After evolving $\vA$, $\vb$, and $\sN$ through 
Eqs. (\ref{eq:Aprime})-(\ref{eq:Nprime}), 
they, along with $\vQ$ as represented
by a single row, are coarse-grained by factors of two (i.e.,
elements summed in the case of
$\vb$ and $\vA$ and averaged in the case of $\vQ$) and the next iteration
proceeds exactly as before. 
All of the problem-specific details go into the
construction of $\vQ(0)$, $\vA(0)$, $\vb(0)$, and $\sN(0)$---after that the 
algorithm proceeds essentially identically for any problem.
After enough iterations the effective data set
becomes small enough to finish the calculation by brute force using the 
analytic integral formula, Eq. (\ref{eq:finalintegral}). 

Note that, while my test problems will be one dimensional, where factors of
two
coarse graining by combining adjacent pixels is the obvious thing to do,
there is no obvious reason not to do this as well in higher dimensions. 
On a cartesian grid we can combine adjacent cells in one direction at a time. 
On a sphere, a hiearchical block of four HEALPixels 
\cite{2002adass..11..107G} can be combined in two steps of pair combinations.
However, it should also be possible to generalize the method to combine 
more than two cells at a time. $\vQ^2$ as discussed above just needs to 
represent the appropriately averaged covariance. 

\subsection{Sparsification}

While the $\vQ^\prime$ cut discussed above limits the range in $\vA$ somewhat,
in practice I find that the evolution of $\vA$ 
produces many small elements that 
do not need to be fully propagated for accuracy and slow down the calculation
significantly.
In \cite{2019PhRvD..99d3538M} I maintained the sparsity of $\vA$ by computing
elements only out to some maximum separation, taken to be a multiple of 
the RG distance scale $\lambda$.
Here I suggest a potentially more 
generally adaptive method, along the lines of the element size
cut discussed above involving $\epsilon_{\vQ^\prime}$. 
The key equation numerically is Eq. (\ref{eq:Aprime}), 
because the matrix products
there dominate the computation time. To control this, I introduce two more 
numerical parameters. When evaluating $\vA\vQ^\prime\vA$, 
I first trim $\vA$ using another threshold parameter, $\epsilon_{\vA}$, 
again basing the cut on the absolute
value of elements relative to the maximum absolute value. To be clear, 
I am not permanently dropping part of the stored, evolving $\vA$, only the 
matrix used to compute $\vA\vQ^\prime\vA$.
I apply another similar cut defined by $\epsilon_{\vA^\prime}$ to 
$\vA^\prime = \vA\vQ^\prime\vA$, before using it to update $\vA$ in each
$\lambda$ step. In practice, for simplicity, I only use
one of these two cuts at a time, finding the $\epsilon_{\vA}$ cut to be
slightly more efficient in my test problems. 

\subsection{Numerical demonstration}

For numerical tests I use one dimensional scenarios
similar to \cite{2019PhRvD..99d3538M}.
I use signal power spectrum $P(k)=A (k/k_p)^\gamma \exp(-k^2)$
with $\gamma=0$ or $-0.5$, where $k$ is measured in units of the data cell
size.  I add unit variance noise to each cell.
I generate mock data with $A_0=1$ and
calculate likelihoods as a function of $A$.
I use pivot $k_p=0.1$ so that the $\gamma=-0.5$ case has both signal and
noise dominated ranges of scales.
To be sure the test covers both 
fine structure and edges, I create
statistically inhomogeneous data sets where the rms 
noise level
in every fourth cell is multiplied by a factor of 10, and the noise in the
last quarter of the data vector is similarly multiplied.

It is more difficult to make a non-trivial test with the innovations in this
paper, because if I assume periodic data with homogeneous noise so that
I can compute the exact likelihood to compare to using FFTs, the obvious
choice of $\vA_\star$ sets $\vA\equiv 0$ so the RG evolution is almost 
trivial. If I also find the maximum likelihood field to use for 
$\vphi_0$, so that $\vb\equiv 0$, it is completely trivial.
For this reason I only do tests with inhomogeneous data in this paper, first
on data sets small enough to compute the exact likelihood by brute force
linear algebra, demonstrating that the RG method works precisely in the
appropriate limit of the numerical parameters, then with
large data sets where the truth is determined by using much better than 
necessary values for numerical parameters. 

After some experimentation, my standard numerical parameter settings are
as follows:
$\vA_\star$ is set to $0.47 N_0^{-1}$, where $N_0$ is the noise power in the 
good part of the data---this sets the accumulated 
${\rm Tr} [\vA\vQ^\prime]$ term in
Eq. (\ref{eq:Nprime}) to approximately zero (the results are 
insensitive to the exact value of $\vA_\star$, as long as it is reasonable).
I specify the number of mid-point method $\lambda$
steps per factor of two coarse
graining by a numerical parameter $N_{d\vQ^\prime}$.
My standard setting is $N_{d\vQ^\prime}=8$ (in an advanced version
of the method, one could try to apply all the usual tricks for solving 
differential equations numerically). I set $\epsilon_{\vQ^\prime}=0.02$,
and $\epsilon_{\vA}=0.0005$. 

\subsubsection{Small problems}

\begin{figure}[ht!]
 \begin{center}
 \includegraphics[scale=1.0]{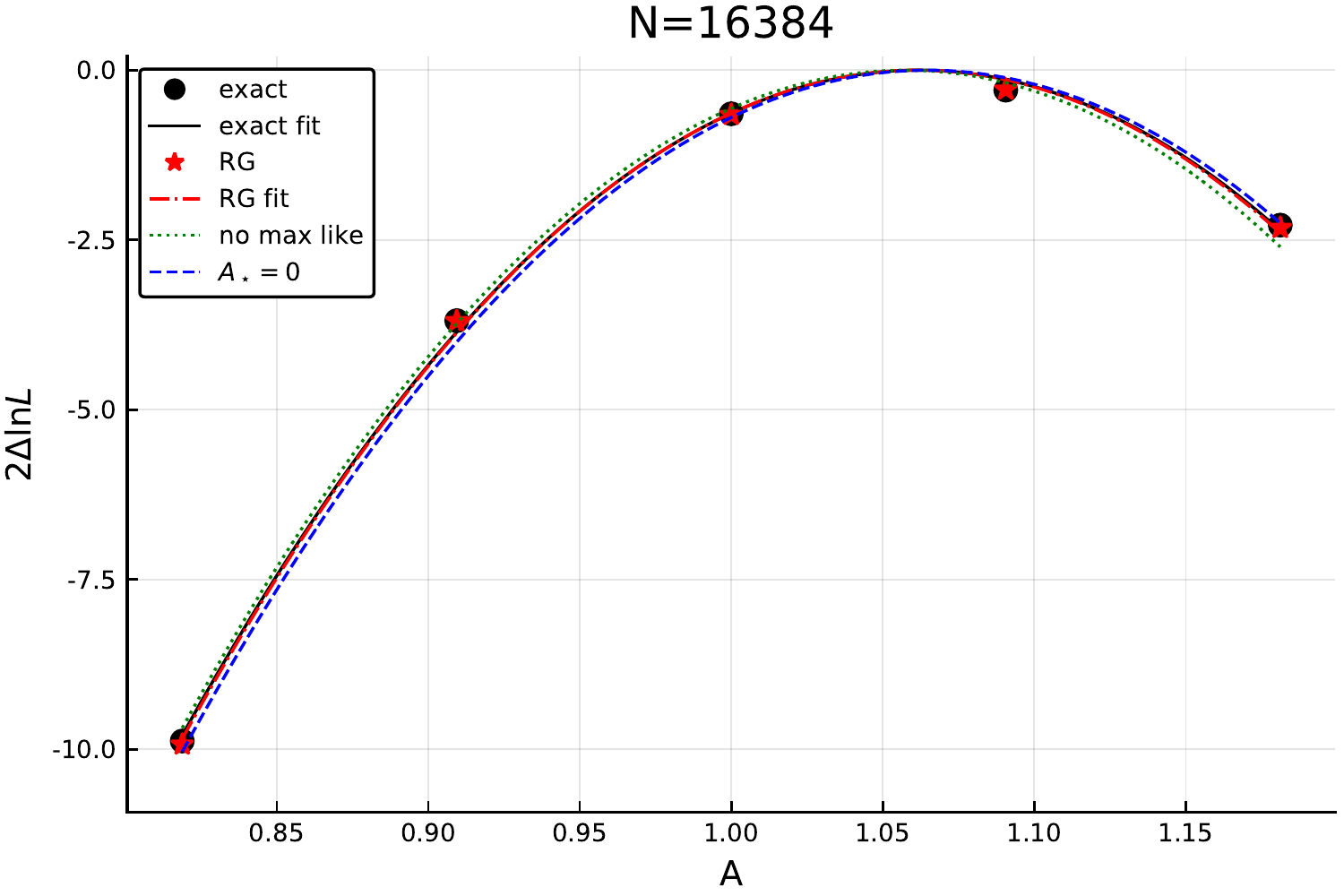}
 \end{center}
 \caption{$N=16384$ test.
The exact likelihood is computed by brute force
linear algebra at five representative values of $A$. To guide the eye, I 
fit a quadratic polynomial to the points, using this to define the 
maximum. I use the RG method to compute the likelihood at the same 
five points,
and similarly plot a quadratic fit representation---the results are 
essentially indistinguishable in this example. For the case with no
maximum likelihood field, i.e., $\vphi_0=0$, and the case with 
$\vA_\star=0$, I plot only the fitted quadratic, to reduce clutter. 
}
 \label{fig:small}
\end{figure}

I first do some tests with $N=16384$, where we can still pretty quickly compute
the exact likelihood by brute force linear algebra, shown in 
Fig. \ref{fig:small}.
The results are good, by construction of course. 
Both using a maximum 
likelihood $\vphi_0$ and using $\vA_\star$ to remove the mean
effect of $\vA$ from the evolution improve the accuracy at fixed parameter
settings, although for these
settings (which were driven by larger data sets) the difference is not 
critical. This example has $\gamma=-0.5$, which is generally a little more
difficult for the algorithm than $\gamma=0$. 

\subsubsection{Large problems}

\begin{figure}[ht!]
 \begin{center}
 \includegraphics[scale=1.0]{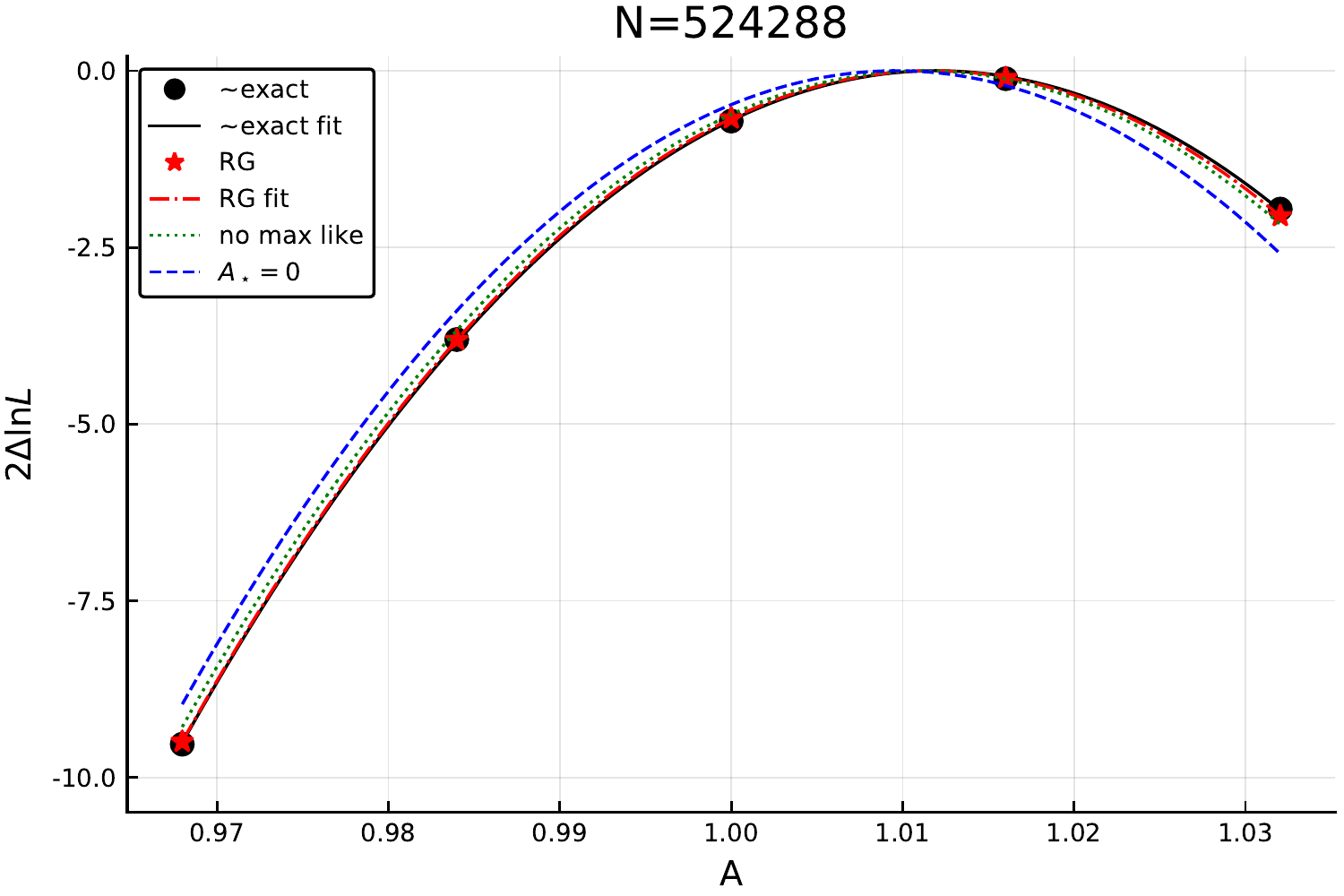}
 \end{center}
 \caption{$N=524288$ test similar to Fig. \ref{fig:small}.
The ``exact'' likelihood is not strictly exact, but
computed for $N_{d\vQ}=25$, $\epsilon_{\vQ^\prime}=0.0005$, and 
$\epsilon_{\vA}=0.0002$, which is perfectly converged at the level of 
differences in this figure. 
}
 \label{fig:large}
\end{figure}

If we are convinced that the algorithm works in the sense of producing 
accurate results in the appropriate limit of numerical parameters, we can
do non-trivial large-scale tests by simply looking for convergence as the
numerical parameters are changed, i.e., we assume that if there is 
convergence it is to the correct result. Figure \ref{fig:large} shows an 
$N=524288$ test, for $\gamma=-0.5$ again. 
The results are again excellent. One might guess based on these figures that
my numerical parameter settings are too conservative, i.e., that I could loosen
them to achieve better speed. This is not actually true---there seems to be
some cancelation of errors that makes the results in these particular 
examples so perfect, and they go
bad very quickly if the parameters are loosened. 

I stop at $N=2^{19}$ for these examples
because careful testing on my laptop becomes tedious beyond this, 
especially running with
extremely conservative parameter settings to be certain of the exact result.
I have run up to two million cells with good looking results. A 
one million cell example runs in two minutes. 
At four million I start to exhaust the memory on my laptop in my current
Julia implementation, although it would be possible to go somewhat further
with more optimization. In any case, it is clear that billion cell data 
sets could be done comfortably on a supercomputer.  

I tried evolving using $Q(\lambda,k)=Q(0,k)e^{-k^2 \lambda^2}$, 
more like in   
\cite{2019PhRvD..99d3538M}, but with a maximum likelihood $\vphi_0$, 
$\vA_\star$, and element size cuts as introduced in this paper, but was unable
to come within a factor of ten of the performance of the pairwise suppression
approach of this paper. 

\section{Discussion}

To summarize, I have suggested the following improvements to the 
basic RG approach of \cite{2019PhRvD..99d3538M}:
\begin{itemize}
\item Integrating out the difference between cells that are to be combined,
rather than small-scale structure more generally, by defining 
$\vQ^\prime$ directly to be proportional to the difference between the current 
and target covariance.
\item Shifting integration variables to integrate around a maximum likelihood
signal field, if available, as $\vphi_0$.
\item Subtracting a statistically homogeneous approximation  
out of the numerically evolving matrix $\vA$, through the 
definition of $\vA_\star$.
\item Cuts on matrix element size, specified by $\epsilon_{\vQ^\prime}$,
$\epsilon_{\vA}$, etc., instead of a simple range cut.
\end{itemize}
The first of these is by far the most important.
In the end it is clear that the algorithm is fast and straightforward 
enough for convenient practical data analysis. 

It was surprising to me that the pair-oriented definition of $\vQ^\prime$
made such a large (factor $\gtrsim 10$) difference in speed.
While the
the principle that if we know which cells we will combine we should focus
on integrating out the difference between them seems good enough to expect 
some improvement, 
I would have been happy with a factor of two. 
It may be that I do not have the best possible implementation of the 
smooth cutoff option. In any case though, it seems like the pair-oriented
approach is the way to go. 

Of course it is only useful to integrate around a maximum likelihood field if 
that field can be found more quickly than the RG analysis could be done without
it. This was the case in my tests, where finding the maximum likelihood field
by conjugate gradient (CG) takes about 5\% of the time in each
likelihood computation. This might not always be the ratio, as my CG 
solution was massively accelerated by being able to multiply by 
things like $\vP$ in Fourier space, including for preconditioning (e.g.,
without preconditioning finding a maximum likelihood field takes longer than
the RG integration without it). If, e.g., the CG had to be done using 
less efficient spherical harmonic transforms, it might be faster not to
use it. 
An interesting possibility is to use the RG method itself to find the
maximum likelihood field. \cite{2019PhRvD..99d3538M} showed how 
to find the data-constrained mean of any function of $\vphi$, with
$\left<\vphi\right>$ itself being the simplest possible version of this. For a
Gaussian problem $\left<\vphi\right>$ is the maximum likelihood field, while 
for a 
non-Gaussian problem it is not but would probably be a better starting 
point than
the maximum likelihood field in that case anyway. Finding $\left<\vphi\right>$ 
can be piggybacked
on a standard likelihood computation with minimal extra cost, but to get 
a speedup in likelihood calculations you would need to feed the result back
into a recalculation. This would only be effective if a useful estimate of 
$\left<\vphi\right>$ could be found with looser numerical settings than would be
required to do the calculation with $\phi_0=0$, which seems quite possible. 
When,
e.g., computing derivatives with respect to parameters,  
we would probably achieve most of the benefit by computing 
$\left<\vphi\right>$ only for the central model (remember that accurate 
results can be achieved for any $\vphi_0$, it is just a question of how tight
numerical settings need to be to do it).

Note that it may not always be beneficial to use $\vA_\star \neq 0$. There is
no cost if all cells in a formal data vector have measurements, i.e., there
are no zeros on the diagonal of
$\vR^t\vN^{-1}\vR$, but if a substantial number of cells 
represent large holes in the
data set or zero padding, so that these elements of $\vA(0)$ can be dropped
from sparse storage, 
setting $\vA_\star \neq 0$ will remove this possibility.
This must be considered on a problem-by-problem basis. 

While my prototype code is already quite fast, at two minutes per likelihood
evaluation per million cells,
there is clearly more room for optimization. Most obviously, I am not taking
advantage of the fact that $\vA$ and $\vQ^\prime$ are symmetric matrices at
all, for no better reason than not knowing canned operations in 
Julia that will do this. Other simple improvements could be tuning of things
like the cuts I've parameterized by $\epsilon_{\vQ^\prime}$, etc.. I kept
these cuts constant for all iterations but this could be wasteful if the 
required
cut value is set by coarser levels of the calculation that do not take 
much total time. A less obvious but I think promising optimization
idea is the following:
The effect of evolving Eq. (\ref{eq:Aprime}) is non-linear in the $\vQ^\prime$
matrix as initial 
changes in $\vA$ are multiplied back together to find the next
step, i.e., we get
products of $\vQ^\prime$ with itself. The required number of steps is surely
set by the products of the largest elements of $\vQ^\prime$---the products
of small elements are perturbatively much smaller. 
This suggests that $\vQ^\prime$
could be split into two or more pieces based on element size. The 
piece(s) with larger elements, which would be very short-range (i.e., 
few elements, i.e., fast to multiply), could be 
evolved first, then longer-range 
the pieces with smaller elements evolved with fewer steps, possibly even one,
because their self-products are negligible. As long as our set of $\vQ^\prime$
steps integrates to
$\vQ_2-\vQ_1$, we are free to choose the details.  

The next step is to implement this for realistic cosmological scenarios. 

\acknowledgments

I thank Zack Slepian and \uros\ Seljak for helpful comments. 
This work was supported by the U.S. Department of Energy, Office of Science, 
Office of High Energy Physics, under Contract No. DE-AC02-05-CH11231. 

\bibliography{cosmo,cosmo_preprints}

\appendix

\section{Alternative approach to integrating out differences between cells
\label{ap:altpairint}}

Before realizing I could define $\vQ^\prime$ by simply differencing the
current and target $\vQ$s, I worked out a 
method for integrating out the difference between cells closer to the
original approach in \cite{2019PhRvD..99d3538M}. 
I include it here to promote
broader understanding of the possibilities.

The RG integration will be controlled by a parameter $\alpha$
which starts at zero and is taken to $\infty$. $\vQ$ and $S_I$ become 
functions of this parameter, i.e.,
\be
\vQ^{-1}(\alpha) \equiv \vP^{-1} +\alpha \vK~,
\ee 
with $\vK$ a fixed matrix to be specified.
Obviously we can suppress fluctuations between cells 1 and 2 by adding a 
term to $S(\phi)$ proportional to $(\phi_1-\phi_2)^2$.  
Repeating this over and over (e.g., $(\phi_3-\phi_4)^2$, etc.)
is equivalent to making $\vK$ the following
block diagonal matrix:
\be
\vK = \left[ \begin{array}{ccc}
\vk & 0 & ... \\
0  & \vk & ...  \\
...  & ... & ...  \\
\end{array} \right] ~,
\ee 
where 
\be
\vk = \left[ \begin{array}{rr}
1 & -1 \\
-1  & 1  \\
\end{array} \right] ~.
\ee
I.e., by dialing $\alpha$ from 0 to $\infty$ in 
$\vQ^{-1}=\vP^{-1}+\alpha \vK$,
we will have effectively integrated out the differences between adjacent 
pairs of cells.
We now have
\be
\vQ^\prime = -\vQ \vK\vQ=-\left(\vP^{-1}+\alpha\vK\right)^{-1} \vK 
\left(\vP^{-1}+\alpha\vK\right)^{-1} ~.
\ee
Unlike in \cite{2019PhRvD..99d3538M}, 
$\vK$ is not exactly translation 
invariant, so we can't simply compute
$\left(\vP^{-1}+\alpha\vK\right)^{-1}$ in Fourier space. 
The structure of $\vQ^\prime$ is the same everywhere, however, up to a 
distinction between odd and even cells, and it is limited to short range,
so we can compute it by brute force
inversion for a limited representative stretch of cells 
and then translate it everywhere.

This approach worked in preliminary tests, 
but not as efficiently as the one in the paper. 

\end{document}

%% file: mydefinitions.tex
\newcommand{\frachalf}{\frac{1}{2}}

\newcommand{\be}{\begin{equation}}
\newcommand{\ee}{\end{equation}}

\newcommand{\sN}{\mathcal{N}}

\newcommand{\vtheta}{\bm{\theta}}

\newcommand{\vk}{\mathbf{k}}

\newcommand{\vb}{\mathbf{b}}

\newcommand{\vmu}{\bm{\mu}}

\newcommand{\vphi}{\bm{\phi}}

\newcommand{\vdelta}{\bm{\delta}}

\newcommand{\vep}{\bm{\epsilon}}
\newcommand{\vo}{\mathbf{o}}

\newcommand{\vC}{\mathbf{C}}

\newcommand{\vQ}{\mathbf{Q}}

\newcommand{\vP}{\mathbf{P}}
\newcommand{\vN}{\mathbf{N}}

\newcommand{\vA}{\mathbf{A}}

\newcommand{\vK}{\mathbf{K}}

\newcommand{\vR}{\mathbf{R}}
\newcommand{\Tr}{\mathrm{Tr}}

\newcommand{\vW}{\mathbf{W}}

\newcommand{\vI}{\mathbf{I}}

\newcommand{\mnras}{{\em Mon. Not. Roy. Astron. Soc. }}

\newcommand{\jcap}{JCAP}
\newcommand{\physrep}{{\em Phys. Rept. }}
\newcommand{\aap}{{\em Astron. Astrophys. }}

\newcommand{\uros}{Uro\v{s}}

\def\prd{{\em Phys. Rev. }{\bf D }}

\def\pvmhid#1{}

\def\afhid#1{}

\def\ashid#1{}